%% file: main.tex
\documentclass[sigconf,authorversion]{acmart}

\usepackage{multirow} 
\usepackage{bm}
\usepackage{float}
\usepackage{subcaption}

\AtBeginDocument{%
  }

\copyrightyear{2025}
\acmYear{2025}
\setcopyright{rightsretained}
\acmConference[RecSys '25]{Proceedings of the Nineteenth ACM Conference on Recommender Systems}{September 22--26, 2025}{Prague, Czech Republic}
\acmBooktitle{Proceedings of the Nineteenth ACM Conference on Recommender Systems (RecSys '25), September 22--26, 2025, Prague, Czech Republic}
\acmDOI{10.1145/3705328.3759333}
\acmISBN{979-8-4007-1364-4/2025/09}

\begin{document}

\title{Benefiting from Negative yet Informative Feedback by Contrasting Opposing Sequential Patterns}

\author{Veronika Ivanova}
\orcid{0009-0002-7080-4648}
\affiliation{%
  \institution{Skolkovo Institute of Science and
Technology}
  \city{Moscow}
  \country{Russia}
}
\affiliation{%
  \institution{Sber AI Lab}
  \city{Moscow}
  \country{Russia}
}
\email{veronika.ivanova88@yandex.ru}

\author{Evgeny Frolov}
\orcid{0000-0003-3679-5311}
\affiliation{%
  \institution{AIRI}
  \city{Moscow}
  \country{Russia}}
\affiliation{%
  \institution{HSE University}
  \city{Moscow}
  \country{Russia}
}
\affiliation{%
  \institution{Skolkovo Institute of Science and
Technology}
  \city{Moscow}
  \country{Russia}
}
\email{frolov@airi.net}

\author{Alexey Vasilev}
\orcid{0009-0007-1415-2004}
\affiliation{%
  \institution{Sber AI Lab}
  \city{Moscow}
  \country{Russia}
}
\affiliation{%
  \institution{HSE University}
  \city{Moscow}
  \country{Russia}
}
\email{alexxl.vasilev@yandex.ru}
\renewcommand{\shortauthors}{Ivanova et al.}

\begin{abstract}
  \input{content/0_abstract}
\end{abstract}

\ccsdesc[500]{Information systems~Recommender systems}

\keywords{Recommender Systems, Sequential Recommendations, Negative Feedback, Contrastive Learning}

\maketitle

\section{Introduction}
\input{content/1_intro}

\section{Related Work}
\input{content/2_related_work}

\section{Proposed Approach}
\input{content/3_approach}

\section{Experimental Settings}
\input{content/4_experimental_settings}

\section{Results and Discussion}

\input{content/5_results}

\vspace{-5pt}
\section{Conclusion}
\input{content/6_conclusion}

\begin{acks}
We sincerely thank Alexandra Laricheva, Sergey Voronin, and Bogdan Monogov for their valuable contributions to this work, including preparing the supplementary materials and enhancing the manuscript’s quality and clarity. The work of A. Vasilev was performed within the framework of the HSE University Basic Research Program.
\end{acks}

\bibliographystyle{ACM-Reference-Format}
\bibliography{content/bibliography}
\end{document}

%% file: content/0_abstract.tex
We consider the task of learning from both positive and negative feedback in a sequential recommendation scenario, as both types of feedback are often present in user interactions. Meanwhile, conventional sequential learning models usually focus on considering and predicting positive interactions, ignoring that reducing items with negative feedback in recommendations improves user satisfaction with the service. Moreover, the negative feedback can potentially provide a useful signal for more accurate identification of true user interests. In this work, we propose to train two transformer encoders on separate positive and negative interaction sequences. We incorporate both types of feedback into the training objective of the sequential recommender using a composite loss function that includes positive and negative cross-entropy as well as a cleverly crafted contrastive term, that helps better modeling opposing patterns. We demonstrate the effectiveness of this approach in terms of increasing true-positive metrics compared to state-of-the-art sequential recommendation methods while reducing the number of wrongly promoted negative items.

%% file: content/1_intro.tex
In the rapidly evolving field of recommender systems, the utilization of negative feedback has emerged as a crucial area of study. While traditional recommendation algorithms have primarily focused on positive interactions, recent research has highlighted the significant potential of incorporating negative feedback to enhance the accuracy and relevance of recommendations, particularly in sequential recommendation scenarios \cite{wang2023learning,pan2023}.

The impact of the negative feedback is significant. 
For instance, there are many ways for users to show their dissatisfaction with a certain song on a music streaming platform: they might skip it, dislike it, or simply not finish listening to it. All of these signs are beneficial to a recommender system since it can learn to recommend fewer songs that resemble disliked ones and more songs that users might like. 
If negative feedback is ignored, the user will again and again receive items that are not relevant at the moment, since the system will not receive new information about interactions, or will receive incorrect data at the input of the model that only considers all interactions as positive.
Apart from providing more information about user feedback that helps to better contrast the true positive preferences, negative feedback naturally extends the notion of learning from shared interests in collaborative filtering. Indeed, if users may share what they like, they may also share their dislikes. Moreover, dislikes alone, if properly accounted for, can already provide enough information for generating relevant recommendations \cite{frolov2016fifty}.

This paper explores the innovative approaches to leveraging negative feedback in sequential prediction tasks for recommender systems. In our work, we obtain separate transformer representations for negative and positive items from the user's history to better distinguish positive and negative signals.


In short, the main contributions of this paper are:
\begin{itemize}
\item We propose a new model architecture to account for negative feedback in sequential recommendations by processing positive and negative user interaction sequences separately.

\item We introduce a new composite three-part loss that makes the model sensitive to negative feedback and helps it to learn better user representations.
\item We train our model using only one type of feedback, either explicit or implicit.
\end{itemize}
We conduct extensive experiments on a diverse range of datasets with varying statistical properties and demonstrate the advantages of our approach over the competing methods.

%% file: content/2_related_work.tex
Leveraging negative feedback in recommender systems is an active area of research. The earlier works rely on non-sequential architectures, such as matrix factorization \cite{wen2019leveraging} and tensor-based collaborative filtering \cite{frolov2016fifty}, while the latest methods are mainly graph-based \cite{seo2022siren, chen2024sigformer} or sequence-based \cite{seshadri2023leveraging, seshadri2024enhancing, wang2023learning, mei2024negative, mei2024hard, gong2022positive, xie2021deep}.

Wen et al. \cite{wen2019leveraging} incorporates implicit negative feedback, such as skips, into existing collaborative filtering algorithms, particularly $BPR$ and $WRMF$. $TC-CML$ \cite{paudel2018loss} models negative user preferences using a metric learning framework. Frolov et al. \cite{frolov2016fifty} treat user feedback as a categorical variable and apply a tensor factorization. 

$SiReN$ \cite{seo2022siren} takes advantage of both high and low ratings via a bipartite graph consisting of GNN, MLP, and attention block. $SIGformer$ \cite{chen2024sigformer} utilizes the transformer architecture with position encodings that capture the spectral properties and path patterns of the signed graph. $DEERS$ \cite{zhao2018recommendations} models the sequential interactions between users and a recommender system as a Markov Decision Process, and uses reinforcement learning to automatically learn the optimal recommendation strategies by incorporating positive and negative feedback.

One way to design a negative feedback informed sequential recommender is to use a transformer encoder with a loss function, constructed to deal with two types of user feedback. The contrastive loss term, in addition to the basic learning objective, is used to leverage negative signals for music recommendations in \cite{seshadri2023leveraging, seshadri2024enhancing}. Gong et al. \cite{gong2022positive} and Xie et al. \cite{xie2021deep} both define implicit negative feedback as cases where an item was shown to the user but not clicked. Additionally, Xie et al. incorporate dislikes as explicit negative feedback signals. Wang et al. \cite{wang2023learning} incorporate explicit and implicit negative user feedback into the training objective using a loss function that optimizes the log-likelihood of not recommending items with negative feedback. Another way to incorporate negative feedback is to replace a proportion of randomly sampled negatives with hard negatives during training \cite{mei2024negative, mei2024hard}, which increases both prediction accuracy and diversity of recommendations. 

Our work focuses on improving the quality of sequential recommenders with negative user responses. The limitation of existing approaches is that they only modify the learning objective \cite{seshadri2023leveraging, seshadri2024enhancing, gong2022positive, xie2021deep, wang2023learning} or use hard negative samples during training \cite{mei2024negative, mei2024hard}, leaving the transformer encoder architecture unchanged. To improve the ability to exploit negative feedback, we propose to train two separate encoders for each type of sequence that distinguish between positive and negative signals. Unlike previous approaches, we employ a cross-entropy-based loss function instead of binary cross-entropy to enhance model performance. Our model is trained exclusively on a single feedback type, explicit or implicit, which simplifies both data collection and processing. To prove the reduction of negative recommendations, we propose a transparent model evaluation: we compute the same conventional metrics, Hit Rate and NDCG, separately for positive and negative ground truth.

%% file: content/3_approach.tex
\begin{figure}
    \centering
        \includegraphics[width=.95\linewidth]{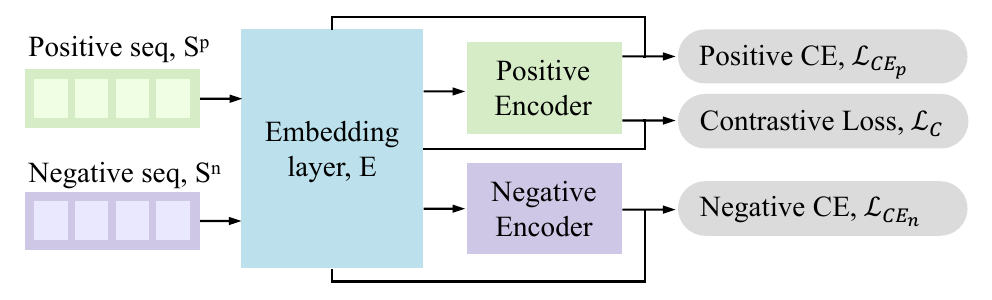}
        \caption{The overall architecture of the proposed \bm{$PNFRec$} model.}
     \label{fig:model}
\end{figure}

\subsection{Problem statement}
The goal of a standard feedback-agnostic sequential recommendation task is to predict the next item $i_{t + 1}$ at time $t + 1$ for each $i_t \in S$, where $S = (i_1, i_2, ..., i_{|S| - 1})$ is the sequence of user interactions, where $1 \leq t \leq |S| - 1$. This approach is not sensitive to the fact that the sequence may contain items with both positive and negative user responses. It also does not penalize the presence of negative items among the recommended items. However, it is important to consider avoiding recommendations that the user dislikes. For example, the proportion of music tracks skipped by a user, indicating a potential dislike, serves as a typical metric of online music recommendations evaluation \cite{brost2019music, germain2013spotify}. Thus, it is not only important to increase the number of items that the user likes, so-called true positives, but also to reduce the number of false positives, recommended items that the user does not like.  

Motivated by this observation, we propose a new approach based on the use of two types of user' feedback: positive and negative responses to items they have interacted with. First, we split the sequence of interactions $S$ into two: $S^p$ and $S^n$, and collect the last $l$ positive and last $l$ negative items from the sequence $S$. Thus, we redefine the problem as predicting the next positive item $i_{t + 1}$ for each $i_t \in S^p$.

\subsection{Model architecture}
Our method is based on the unidirectional transformer
architecture, specifically the $SASRec$ \cite{kang2018self} model. The overall architecture of the proposed PNFRec
model is demonstrated on Figure 1. The key feature of our proposed method is that we adapt the idea of learning signals of different natures of user behavior separately \cite{baikalov2024end}. In particular, we use two transformer encoders to process positive $S^p$ and negative $S^n$ user interaction sequences to learn the connection between successive positive $i_t \in S^p$ and negative $j_t \in S^n$ items separately. 

\subsubsection{Embedding layers}
We store learned embeddings in a lookup table $e_t \in E$ of size $N \times d$, where $N$ is the number of items and $d$ is the dimensionality of the embedding. Each item has a single embedding, regardless of whether the interaction received positive or negative feedback.

We include information about the position of each item in the sequence following existing approaches \cite{kang2018self, xie2022contrastive}. We add a learnable positional embedding $P$ of size $l \times d$, where $l$ is the maximum length of the sequence. Positive $S^p$ and negative $S^n$ sequences are modeled in a way that each sequence is encoded independently with its own positional embedding.

\subsubsection{Backbone model}
We employ two standard transformer encoders to learn session-level information for positive $S^p$ and negative $S^n$ sequences. During training, both encoders process their respective sequences. At inference, only the positive encoder is used to predict the next positive item. The negative encoder influences training, since each item has a single embedding from the embedding layer, regardless of feedback type. This model is based entirely on self-attention mechanisms, using multiple self-attention layers along with position-wise feedforward layers to capture contextual information from sequential inputs.

\subsubsection{Inference}
Our task is to predict the next positive item, so to get the predictions we pass the sequence of positive items from the user's history through the positive encoder only.

\subsection{Loss functions}
To jointly train two encoders, we use the loss function for the sequential task, specifically cross-entropy, as it has proved better performance compared to binary cross-entropy loss \cite{klenitskiy2023turning, mezentsev2024scalable}. To train the positive encoder we use the positive CE $\mathcal{L}_{CE_p}$ with the next predicted positive items $i_{t + 1}$ for each $i_t \in S^p$, the inputs of the negative CE loss function $\mathcal{L}_{CE_n}$ are the next predicted negative items $j_{t + 1}$ for each $j_t \in S^n$.

Additionally, we use the contrastive loss $\mathcal{L}_c$ between the normalized embeddings of the next positive item and negative items from the negative sequence of the corresponding user. 

\begin{equation*}\mathcal{L}_c = - \sum_{u \in U} \sum_{i_t \in S^p} log \cfrac{exp(f(\hat{i}_t, i_{t+1}))}{exp(f(\hat{i}_t, i_{t+1})) + \sum_{j \in S^n} exp(f(\hat{i}_t, j))}
\end{equation*}

where $u$ is the user from the set of all users $U$, $\hat{i}_t$ is the representation of $i_t \in S^p$, $f$ is cosine similarity. 

The overall objective is as follows:
\begin{equation*}\mathcal{L} = \mathcal{L}_{CE_p} + \alpha \mathcal{L}_{CE_n} + \beta \mathcal{L}_c
\end{equation*}

where $\alpha$ and $\beta$ are model hyperparameters that determine the influence of each objective on the overall model objective.

%% file: content/4_experimental_settings.tex
\subsection{Datasets}
We conduct experiments on four datasets suitable for our task with a proper sequential structure \cite{klenitskiy2024does}. MovieLens-1m and MovieLens-20m are two versions of the popular movie recommendation dataset \cite{harper2015movielens}. Amazon Toys\&Games is an e-commerce review dataset collected from Amazon.com \cite{mcauley2015image}. In addition to popular academic benchmarks, we add Kion -- the recently released industrial dataset for recommending movies and TV shows with implicit feedback \cite{petrov2022mts}.

For each dataset, we filter out users and items with less than 5 interactions. For our task, we choose the threshold of negative feedback to be the median of the ratings in the explicit datasets. For the implicit Kion dataset, we choose the negative feedback threshold from the distribution of movie watching percentages. The chosen negative feedback threshold is 15, which corresponds to the proportion of negative feedback interactions in the other datasets. The final statistics of the datasets are shown in the table ~\ref{datasets}.

\subsection{Models}
Our model $PNFRec$ incorporates two transformer encoders in its architecture and is trained with a composite learning objective that includes positive and negative CE as well as contrastive loss. To prove the contribution of each term of the loss function, we investigate the quality of $PNFRec$ trained with two components: either using positive and negative CE or positive CE and contrastive loss. We aim to demonstrate that our model and loss function design can increase the true positives among the recommendations as well as reduce the presence of false positives.

\subsubsection{Baseline models}
Existing models for incorporating negative feedback \cite{seshadri2023leveraging, seshadri2024enhancing, wang2023learning, mei2024negative, mei2024hard} are based on the $SASRec$ architecture; therefore, we compare our results to the following baselines: 
\begin{itemize}
    \item $SASRec_p$, trained on positive interactions only;
    \item $SASRec$, trained on the user's entire interaction history;
    \item $SASRec_c$, trained on the loss objective with the contrastive term introduced in \cite{seshadri2024enhancing}.
\end{itemize}

Unlike the original $SASRec$ implementation, which used binary cross-entropy loss, we use full cross-entropy because this approach has been shown to provide a state-of-the-art quality across a wide range of datasets \cite{klenitskiy2023turning,mezentsev2024scalable}.
During model inference, we pass the sequence containing the items with feedback on which the model was trained, namely the sequence of positive items for the first baseline model and the entire sequence for the latter models.

\subsubsection{Hyperparameters selection}
For the Amazon and Kion datasets, we choose a maximum sequence length of 50 for $SASRec$ trained on the full sequence and the baseline model with contrastive loss \cite{seshadri2024enhancing} and 25 for $SASRec$ trained on the positive sequence. For the MovieLens datasets, the sequence length is doubled because the number of items per user is higher than for the other datasets. For our $PNFRec$ model, we also use a negative input sequence with the same maximum length as the positive sequence. 

We tune the hidden size, the number of self-attention blocks, and the learning rate for all sequential models, and use 1 attention head. We tune the optimal $\alpha, \beta \in [0, 0.05, ... 1]$ for $PNFRec$, we also tune the contrastive loss coefficient for the baseline model $SASRec_c$ with the same range of values as $\beta$. The choice of the value range of $\alpha$ is conditioned by the ratio of $\mathcal{L}_{CE_p}$ to $\mathcal{L}_{CE_n}$, we found that $\beta > 1$ negatively affects the quality, so we only considered $\beta \in [0, ... 1]$ during the tuning. Tuning the full set of hyperparameters for our $PNFRec$ model with three loss terms is computationally expensive, so we took the best hyperparameters for $SASRec_p$ and sequentially tuned the coefficients $\alpha, \beta$ while keeping the remaining hyperparameters fixed. We verified that this incremental tuning approach yields metric values equivalent to those obtained by tuning the full hyperparameter set on the smallest ML-1M dataset.

For validation, we use only users with positive ground truth for early stopping, measure $NDCG_p@10$, and stop training if the validation metric did not improve for 10 epochs (patience parameter).
\begin{table*}[!ht]\centering
\caption{Experimental datasets.}
\small
\begin{tabular}{l r r r r r r r} 
 \hline
 \textbf{Dataset} & \textbf{Users} & \textbf{Items} & \textbf{Actions} & \textbf{Feedback Th.} & \textbf{\% Neg. Feedback} & \textbf{Avg. actions/user} & \textbf{Density} \\
 \hline
 ML-1M & 6,040 & 3,412 & 999,611 & 4 & 42\% & 165 & 4.84\%\\ 
 ML-20M & 138,493 & 18,345 & 19,984,024 & 4 & 50\% & 146 & 0.79\%\\ 
 Toys\&Games & 19,412 & 11,924 & 167,597 & 5 & 38\% & 9 & 0.08\%\\ 
 Kion & 302,256 & 10,348 & 4,279,866 & 15 & 40\% & 14 & 0.13\%\\ 
 \hline
\end{tabular}

\label{datasets}
\end{table*}

\begin{figure*}[h!]
    \centering
    \begin{subfigure}[t]{0.45\textwidth}
        \centering
        \includegraphics[width=.85\linewidth]{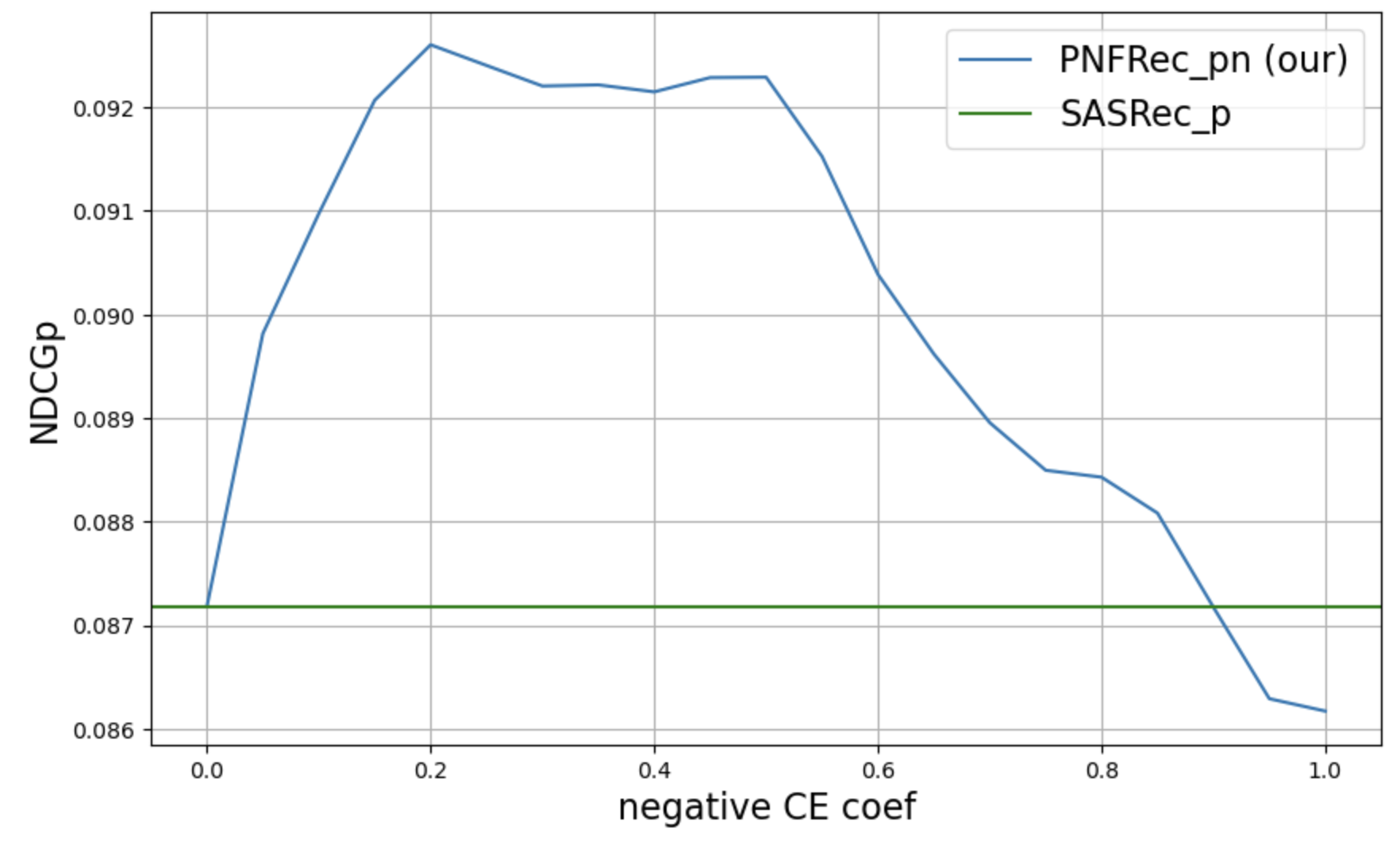}
        \caption{\bm{$NDCG_p@10$} for different negative CE coefficients \bm{$\alpha$}.}
    \end{subfigure}%
    ~ 
    \begin{subfigure}[t]{0.45\textwidth}
        \centering
        \includegraphics[width=.85\linewidth]{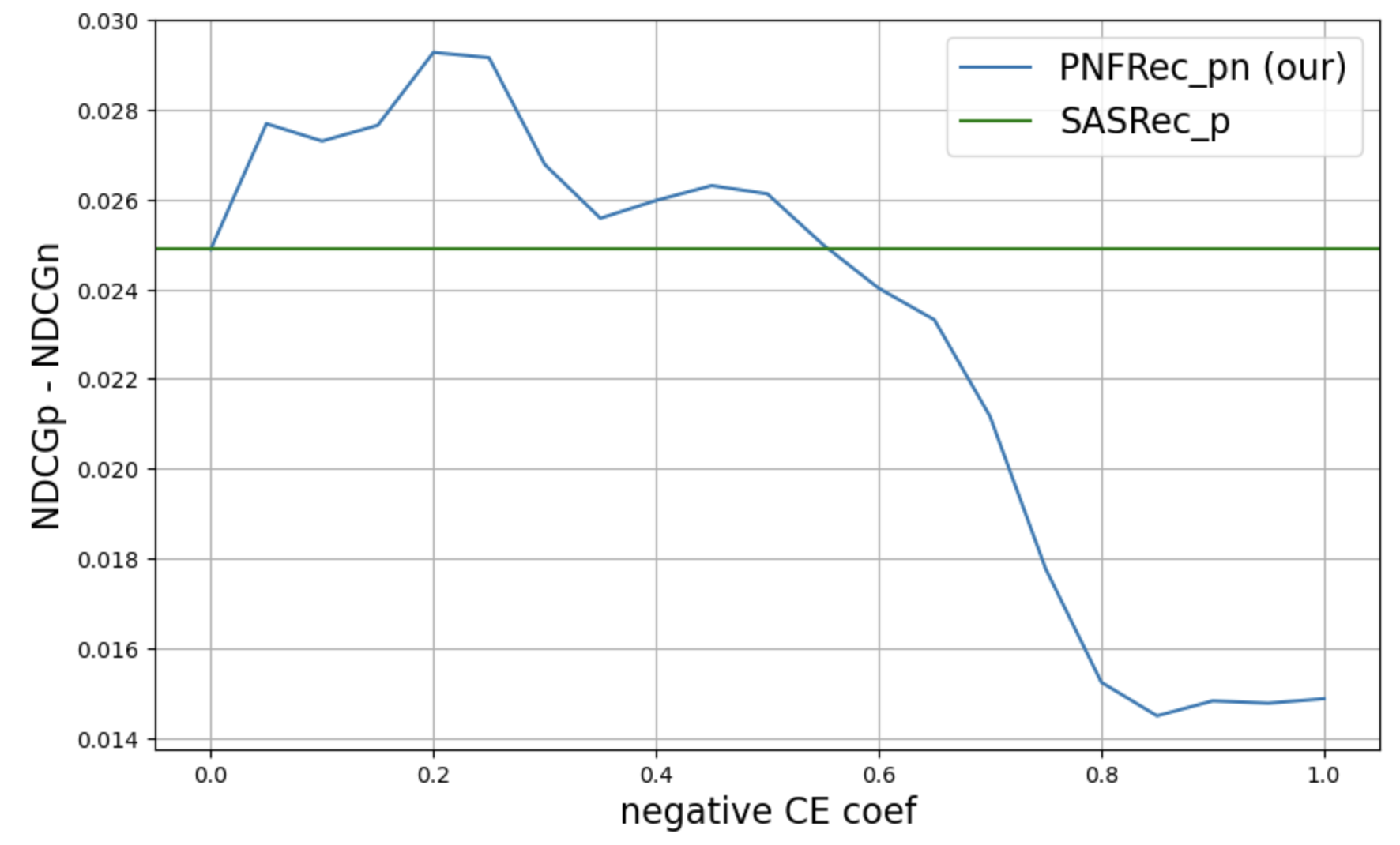}
        \caption{\bm{$\Delta NDCG@10$} for different negative CE coefficients \bm{$\alpha$}.}
    \end{subfigure}
    \caption{Dependence of \bm{$NDCG_p@10$} and \bm{$\Delta NDCG@10$} on \bm{$\alpha$} with fixed hyperparameters for \bm{$PNFRec_{pn}$} for ML-1M.}
     \label{fig:contrastive_coef}
\end{figure*}

\begin{table*}[!ht]
\centering
\caption{Overall Performance Comparison. The best result for \bm{$\Delta HR@10$} and \bm{$\Delta NDCG@10$} in each row is highlighted in bold, the second-best result is underlined. The relative increase compares the best result of our \bm{$PNFRec-based$} models against \bm{$SASRec-based$} baselines. All metrics are averaged over 5 independent runs.}

\label{tab:perf_comp}
   \begin{tabular}{|l| l |l l l|l l l|l|}
    \hline
    Dataset & Metric & $PNFRec (our)$ & \textbf{$PNFRec_{pn} (our)$} & \textbf{$PNFRec_{pc} (our)$} & \textbf{$SASRec_p$} & \textbf{$SASRec_c$} & \textbf{$SASRec$} & \% increase\\ 
    \hline
      \multirow{5}{*}{ML-1M} & $HR_p@10$ & 0.1716	& 0.1677	& 0.1612 & 0.1587	& 0.1638	& 0.1683 & +2\%\\
    
    & $NDCG_p@10$ & 0.1041&	0.1024&	0.0921&	0.0935&	0.0892&	0.0927 & +11\%\\
    & $\Delta HR@10$ & 0.0724&	0.0644&	\textbf{0.0760}&	\underline{0.0726}&	-0.0213&	-0.0160 & +5\%\\
    & $\Delta NDCG@10$ & \textbf{0.0484}&	\underline{0.0467}&	0.0435&	0.0449&	-0.0174&	-0.0116& +8\%
 \\\hline
 
        \multirow{5}{*}{ML-20M} & $HR_p@10$ & 0.0933&	0.0930&	0.0920 &0.0873&	0.0893&	0.0902& +3\%\\
    
    & $NDCG_p@10$ & 0.0541&	0.0544&	0.0538&	0.0511&	0.0541&	0.0545& 0\%\\
    & $\Delta HR@10$ & \textbf{0.0288}&	0.0265&	\underline{0.0278}&	0.0276&	0.0016&	0.0043& +4\%\\
    & $\Delta NDCG@10$ & 0.0168&	0.0162&	\textbf{0.0176}&	\underline{0.0174}&	0.0007& 0.0028& +1\%\\\hline
    
    \multirow{5}{*}{Toys\&Games} & $HR_p@10$ & 0.1109&	0.1119&	0.1008&	0.0998&	0.1279&	0.1275& -12\%\\
    
    & $NDCG_p@10$ & 0.0731&	0.0743&	0.0656&	0.0661&	0.0795&	0.0804& -8\%\\
    & $\Delta HR@10$ & \underline{0.0419}&	\textbf{0.0429}&	0.0364&	0.0354&-0.0284&	-0.0425& +21\%\\
    & $\Delta NDCG@10$ & \underline{0.0377}&	\textbf{0.0385}&	0.0295&	0.0290&	-0.0191&	-0.0256& +33\%\\\hline
    
    \multirow{5}{*}{Kion} & $HR_p@10$ & 0.2157&	0.2158&	0.2128&	0.2135&	0.2182&	0.2180& -1\%\\
    
    & $NDCG_p@10$ & 0.1310&	0.1311&	0.1292&0.1300	&	0.1308&	0.1316& 0\%\\
    & $\Delta HR@10$ & \textbf{0.1036}&	\underline{0.1034}&	0.1026&	0.1024&	0.0727&	0.0719& +1\%\\
    & $\Delta NDCG@10$ & \underline{0.0700}&	\textbf{0.0701}&	0.0682&	0.0689&	0.0506&	0.0521& +2\%\\\hline
\end{tabular}
\end{table*}

\subsection{Evaluation}
The most common evaluation strategy for the next item prediction task is the leave-one-out strategy \cite{kang2018self, sun2019bert4rec}.  However, it can lead to data leakage and may not accurately reflect the performance of recommendation models in production. To preserve the sequential aspect of the recommendation task and avoid data leakage, we combine global temporal and leave-one-out splitting. 

For each dataset, we select the global temporal boundary corresponding to 90\% of the interactions and take all previous interactions into the training subset. We divide the subsequent interactions equally by users into the test and validation subsets. For the test and validation sets, we consider the last interaction of the user as ground truth and all previous interactions as an input sequence.

Traditional evaluation approaches only assess true-positive metrics such as $HR@k$, $MRR@k$, and $NDCG@k$ \cite{mena2020agreement}, which are blind to the presence of false positives among the recommended items, specifically items with negative responses in the ground truth. We propose to split the set of users' last interactions into two subsets: items with positive feedback and items with negative feedback, and compute metrics on them separately. So we have $NDCG_p@k$ and $NDCG_n@k$, calculated on positive and negative items from the ground truth, and $\Delta NDCG@k = NDCG_p@k - NDCG_n@k$. Our goal is to maximize $NDCG_p@k$ and $\Delta NDCG@k$. We emphasize the importance of evaluating $NDCG_n@k$, as two algorithms may demonstrate comparable performance on $NDCG_p@k$, yet one of them would provide the user with fewer irrelevant recommendations. 
We used the $RePlay$ framework \cite{vasilev2024replay} to compute the metrics.
The code for our experiments is publicly available \footnote{\url{https://github.com/Veronika-Ivanova/pnfrec}}.

%% file: content/5_results.tex
Table ~\ref{tab:perf_comp} shows the results for our $PNFRec$ model, its reduced versions $PNFRec_{pn}$ with \(\mathcal{L}_{pn} = \mathcal{L}_{CE_p} + \alpha \mathcal{L}_{CE_n}\), and $PNFRec_{pc}$ with \(\mathcal{L}_{pc} = \mathcal{L}_{CE_p} + \beta \mathcal{L}_c\), compared to baseline with contrastive loss $SASRec_{c}$ \cite{seshadri2024enhancing}, $SASRec_{p}$ trained on positive sequences and $SASRec$ trained on the whole sequence. 

We find that our composite loss term can increase the difference between metrics calculated on positive and negative ground truth: $\Delta HR@10$ and $\Delta NDCG@10$ across datasets, while maintaining or even improving the true-positive metrics: $HR_p@10$ and $NDCG_p@10$ for the majority of datasets.

We observe that our proposed $PNFRec$ outperforms other models on the densest ML-1M and ML-20M datasets. The negative CE loss component helps improve $HR_p@10$ and $NDCG_p@10$, while the contrastive component mainly affects $\Delta HR@10$ and $\Delta NDCG@10$. For Amazon Toys\&Games $PNFRec$ discriminates between positive and negative items better than the baseline models, as we can see from the $\Delta HR@10$ and $\Delta NDCG@10$ metrics. As for the implicit Kion dataset, $PNFRec$ and $PNFRec_{pn}$ show a better ability to reduce negatives in the recommendations with $HR_p@10$ and $NDCG_p@10$ no worse than for the baseline models. 
Additionally, $PNFRec$ requires less training time than $SASRec_{c}$ across all datasets.

In general, the $SASRec_{c}$ and $SASRec$ models with inference on the whole input sequence are not able to distinguish positive and negative items on explicit feedback datasets, as we can see from the near-zero $\Delta HR@10$ and $\Delta NDCG@10$. The Kion dataset contains implicit feedback, in particular the percentage of movies watched, which is naturally more noisy than explicit ratings, making it difficult to determine the boundary between positive and negative feedback.

Figure ~\ref{fig:contrastive_coef} shows the dependence of $NDCG_p@10$ and $\Delta NDCG@10$ on $\alpha$ for $PNFRec_{pn}$ with the other best hyperparameters chosen during validation. The negative CE loss helps to reduce the negatives in the recommendations when $\alpha < 0.55$ while increasing the positives. The optimal range of the negative CE loss coefficient $\alpha \in [0.1, 0.35]$.

%% file: content/6_conclusion.tex
The paper presents our model PNFRec, which incorporates negative feedback in the sequential recommender with the complex loss objective of positive and negative cross-entropy and contrastive loss terms. We showed a noticeable improvement in quality in terms of reducing negative items and increasing positives among the recommendations on four datasets from different domains. 
Our future work includes enhancing the PNFRec model to handle cases where users have only negative interactions and improving overall performance.